\documentclass[12pt,preprint]{aastex}

\usepackage{graphicx}
\usepackage{natbib}

\usepackage{amssymb}
\usepackage[figuresright]{rotating}

\newcommand{\gapprox}{$\stackrel {>}{_{\sim}}$}   
\newcommand{\lapprox}{$\stackrel {<}{_{\sim}}$}

 \slugcomment{To appear in Ap. J.}

\begin{document}

\title{On the nature of the EXor accretion events:
an unfrequent manifestation of a common phenomenology ?
\thanks{Based on observations collected at AZT-24 telescope
(Campo Imperatore, Italy), AZT-8 (Crimea, Ukraine), and LX-200 (St.Petersburg, Russia)}}

\author{D.Lorenzetti\altaffilmark{1},
S.Antoniucci\altaffilmark{1},
T.Giannini\altaffilmark{1},
G.Li Causi\altaffilmark{1},
P.Ventura\altaffilmark{1},
A.A.Arkharov\altaffilmark{2},
E.N.Kopatskaya\altaffilmark{3},
V.M.Larionov\altaffilmark{2,3,4},
A.Di Paola\altaffilmark{1},
and
B.Nisini\altaffilmark{1}.
}
\altaffiltext{1}{INAF - Osservatorio Astronomico di Roma, via
Frascati 33, 00040 Monte Porzio, Italy, dario.lorenzetti,
simone.antoniucci, teresa.giannini, gianluca.licausi,
andrea.dipaola, brunella.nisini, paolo.ventura@oa-roma.inaf.it}
\altaffiltext{2}{Central Astronomical Observatory of Pulkovo,
Pulkovskoe shosse 65, 196140 St.Petersburg, Russia,
arkadi@arharov.ru}

\altaffiltext{3}{Astronomical Institute of St.Petersburg
University, Russia, vlar2@yandex.ru, enik1346@rambler.ru}
\altaffiltext{4}{Isaac Newton Institute of Chile, St.Petersburg
branch}

\begin{abstract}
We present the results of a comparison between classical and newly identified EXor based on
literature data and aimed at recognizing possible differences or similarities of both categories.
Optical and near-IR two-color diagrams, modalities of fluctuations, and derived values of the mass accretion rates
are indicative of strong similarities between the two samples. We demonstrate
how the difference between the outburst and the quiescence spectral energy distribution
of all the EXor can be well fitted with a single blackbody, as if an additional thermal component appears during the outbursting phase. Temperatures of this additional component span between 1000 and 4500 K, while the radii of the
emitting regions (assumed to be a uniform disk) span between 0.01 and 0.1 AU, sizes typical of the inner portions of the
circumstellar disk. Spots persisting up to 50\% of the outburst duration,
not exceeding the 10\% of the stellar surface, and with temperatures compatible with the EXor mass accretion rates,
are able to account for both the appearance of the additional thermal component and the dust sublimation
in the inner structures of the disk.
We also compare the EXor events with the most significant color and magnitude fluctuations of active T Tauri stars
finding that ({\it i}) burst accretion phenomena should also be important for this latter class; ({\it ii})
EXor events could be more frequent then those accidentally discovered.
Remarkable is the case of the source V2493 Cyg,
a T Tauri star recently identified as a strong outbursting object: new optical and near-IR photometric and spectroscopic data are presented
trying to clarify its EXor or FUor nature.
\end{abstract}


\section{Introduction}

Young stellar objects (YSO's) from low to intermediate mass (0.5-8 $M_{\odot}$), accumulate
a large part of their final mass during the so-called main accretion phase (lasting about
10$^5$ yrs). After this period they continue to accrete, although at lower rates. The
material repeatedly falls from the inner envelope onto their circumstellar disk, then
crosses the viscous part of the disk itself and finally lands onto the star surface through
the magnetic interconnection lines (Shu et al. 1994).
This evolutionary stage is crucial for the comprehension of the star formation processes,
because during this phase the accretion process halts,
determining both the consequent evolution of the protostar and the Initial Mass Function (IMF) slope.

Disk accretion phenomena are characterized by intermittent outbursts (usually
detected in the optical and near-IR bands) due to the sudden increase of the mass
accretion rate (by orders of magnitude, Hartmann \& Kenyon 1985). More recently, D'Angelo \& Spruit (2010)
presented a new work on the instability that can lead to episodic outbursts; they provided quantitative
predictions for the episodic accretion on magnetized stars and
indication that the cycle time of the bursts increases with a decreasing accretion rate.
Such events are typical of many (if not all) YSO's. The outbursts of larger intensity (\gapprox~4 mag)
are classified into two major classes: ({\it i}) EXor events (Herbig 1989) lasting one year or less,
with a recurrence time of months to years, characterized by emission line spectra;
({\it ii}) FUor events (Hartmann \& Kenyon 1985) of longer duration (\gapprox~tens of years)
with spectra dominated by absorption lines.

However, it should be noted that an irregular photometric variability attributable to disk accretion
is a defining feature of all the classical (i.e. most
active) T Tauri stars (CTTS) and YSO's, although at lower levels of amplitude (typically 0.2-1 mag).
Consequently, the optical/near-IR intensity of most young stars tends to present random variations, and the EXor/FUor
events could plausibly represent just extreme and incidentally observed cases of a widespread phenomenum.

Currently, doubts exist about the classification as EXor or FUor of some recently
detected outbursts (see e.g. the case of V2493 Cyg - Miller et al 2011; Semkov \& Peneva 2010;
K\'{o}sp\'{a}l et al. 2011). Sometimes even the origin of
the observed variability is matter of debate (see e.g. the case of GM Cep - Sicilia-Aguilar
et al. 2008; Xiao et al. 2010). These dubious circumstances arise from observations often relying
on a single event and not on a long lasting photometric and/or spectroscopic monitoring. Remarkably, a single outburst does not allow us to ascertain if the object
will remain or not in its higher state; analogously, from a single fading, we cannot decide
whether we are looking at a typically bright object subjected to repetitive obscurations or,
conversely, at a typically quiescent object that undergoes recurrent outbursts.

In the following, we will concentrate on the EXor class, whose typical phenomenology
(outbursts, fading, recurrent activity) occurs on time-scales which are more easily accessible than
those typical of FUor.
So far, about 20 EXor systems are known which can be grouped into two sub-classes: the classical EXor
(Herbig 1989, 2008), and the new identifications (see Sect.2 for the list of references). Hereafter,
for simplicity, we will refer to these classes as classical and newest, respectively. Historically,
the first group was identified in the visual band and, consequently, these accretion-disk systems
are essentially unobscured (i.e. without significant envelopes). Later, the increasing
availability of near-IR facilities quite naturally favored the identification of
more embedded eruptive variables, typically associated with an optical-IR nebula (second sub-class).
The membership to this latter class often relies on sporadic
events and does not stem from a comprehensive analysis aimed at
checking if the object presents all the properties typical of the classical
prototypes (i.e. repetitive outbursts, rapid brightening and slower fading, colors pre-and post-outburst).
In principle, there is no physical reason preventing the EXor phenomenum from occurring also
during the more embedded (i.e. earlier) phase. In this scenario, the classical EXor, having already emerged
from their dust cocoons, might be associated with later phases of the pre-main sequence evolution.

Here, to better clarify the EXor phenomenological framework we want: ({\it i})
to recognize possible similarities (or differences) between the two outbursting classes in terms of
near-IR colors, shape of their spectral energy distribution (SED), and mass accretion rate;
({\it ii}) to compare the EXor variability with the more irregular one which is exhibited by a sample of CTTS
observed in different epochs, searching for any common modality.
In Sect.2 we present the investigated sample. Sect.3 and 4 are dedicated to analyze
our sample and to compare it with other CTTS, respectively. Our optical and near-IR monitoring of the outbursting source
V2493 Cyg is commented in Sect.5. Concluding remarks are given in Sect.6.

\section{The Sample}

The sample is composed by the classical and newest EXor that are listed in the upper and lower part of Table~\ref{sample:tab}, respectively.
Here, the classical sources are identified with their commonly used names, while for some of the newest objects, the adopted names are those given in the General Catalogue of Variable Stars (GCVS - Kazarovets, Reipurth, \& Samus (2011). Indeed,
different identification can be found in the recent literature: HA11 corresponds here to V1180 Cas, CTF93 216-2 to V2775 Ori, ISOChaI 192 to GM Cha, HBC 722 to V2493 Cyg, respectively. Each source is coded with an alphabetic letter (column 2), which will be used in the relevant figures. The range of the A$_V$ parameter, as found in the literature, is given in column 5.
The bolometric luminosity values (listed in column 6) refer sometimes to different parameters (photometric/accretion luminosity), opportunely indicated. In columns 7 and 8 a range of variability is given by listing the minimum
and maximum V-band magnitude ever detected. Relevant references for the given parameters are provided in column 9.

The source PV Cep represents a fairly uncertain case: albeit included in the original Herbig's list, it is quite embedded, luminous, associated to an optical/near-IR nebula, therefore its nature is quite controversial (Kun et al. 2011a;
Lorenzetti et al. 2011b). A comparative study of sample sources will also allow us to better ascertain the
membership of the new entries.\

\section{Analysis and discussion}

This Section will mainly deal with SED's and IR
colors associated with different phases
of both classical and newest EXor.
It is worthwhile to anticipate that, as expected, much more extensive data exist for the newest sources.
Indeed, their eruptive events have been sampled in a greater number of spectral bands (mainly IR) than the classical ones,
whose light-curves were obtained from aperture photometry, typically performed in just one optical band (usually V or photographic).
Due to observational constraints, this study has to rely on a still sparse and incomplete database.
Nevertheless, the increasing number of monitoring efforts allows us to confirm (or not) the
conclusions outlined here.

\subsection{The near-IR two-colors diagram}

\begin{figure}
\includegraphics[angle=0,width=15cm] {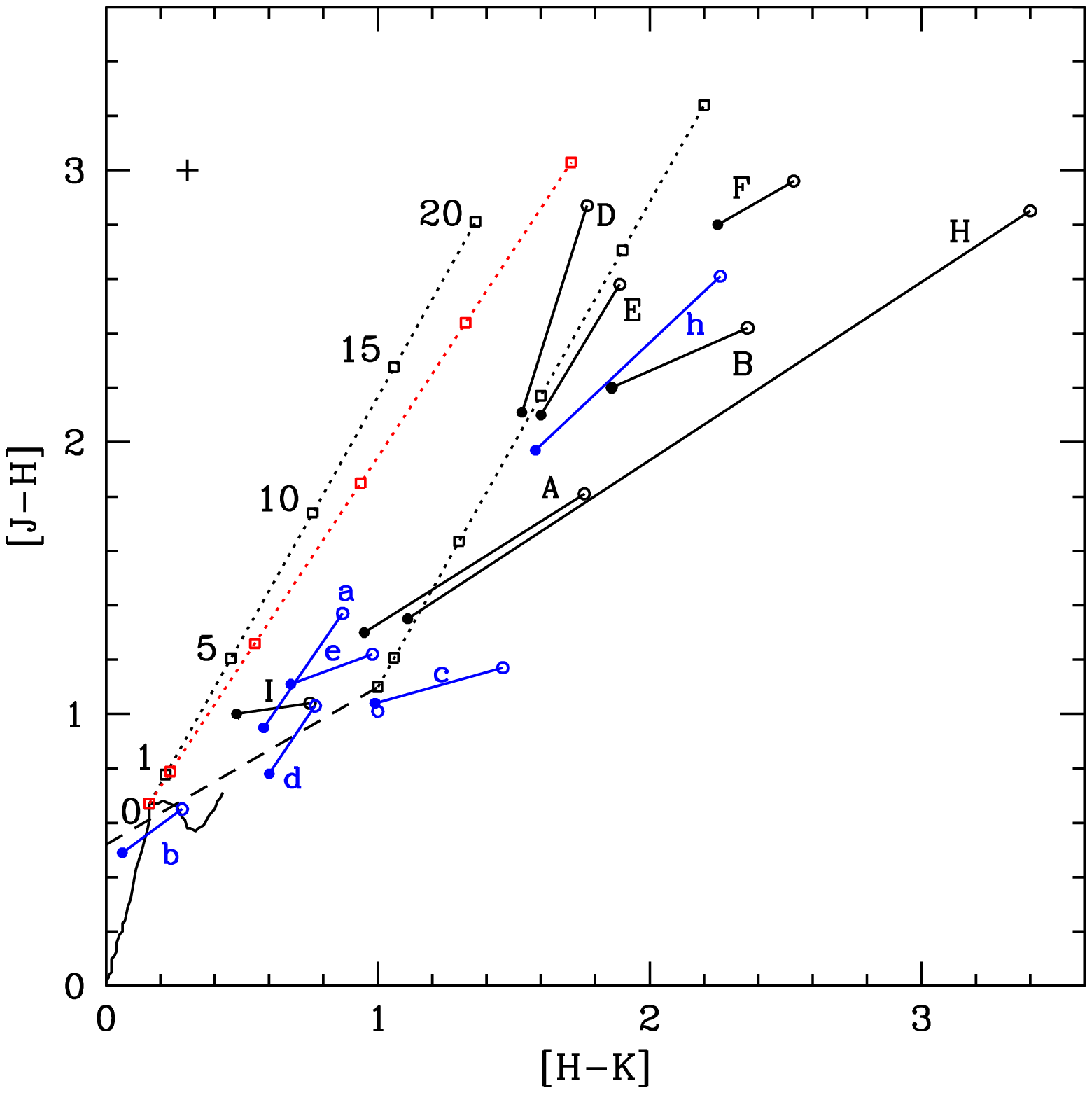}
   \caption{Near-IR two-colors diagram of EXor in different epochs.
   The solid line marks the unreddened main sequence, whereas the dashed
   one is the locus of the T Tauri stars (Meyer et al. 1997). The black and red
   dotted lines represent the reddening law by Rieke \& Lebofsky (1985)
   and Cardelli et al. (1989), respectively, where different values of A$_V$ are indicated by open
   squares. Sources are alphabetically coded as in Table 1. Solid (open) black circles
   identify the new targets during their outburst (quiescent) phase.
   The same is true for the classical EXor stars depicted in blue. These latter evidently overlap much
   better with the typical locus of T Tauri stars. A cross in the upper left corner indicates the typical error.
   \label{colcol_nir:fig}}
\end{figure}

For comparison purposes, in Figure~\ref{colcol_nir:fig} both the classical sources (depicted in blue) and
the newest ones (in black) are displayed into the same two-colors plot [J-H],[H-K]. Magnitude values are taken from the literature data given in Table~\ref{fluxes:tab} and references therein. Source position is indicated by circles: the solid (open) circles indicate the color during the outburst (quiescent)
state. The classical EXor (with the exception of PV Cep) are all located close to the locus of unobscured CTTS and present
color variations that systematically deviate from the extinction law: passing from quiescence to outburst, they
become substantially bluer, but show a [J-H] color that is larger than the one predicted on the basis of a pure extinction (Lorenzetti et al. 2007).
Noticeably, the same behavior is shown also by the newest objects (with the possible exception of V1647 Ori and V2775 Ori),
even if two main differences are recognizable with respect to
the classical ones: ({\it i})  the newest objects occupy a redder
portion of the plot and ({\it ii}) their color fluctuations are larger. The first occurrence is an obvious consequence
of their location within more obscured regions (see the A$_V$ values given in Table~\ref{sample:tab}), as is often
witnessed by the presence of an associated nebula; the second one comes from the fact that the light-curve of the newest sources was completely followed in the near-IR (thus providing the maximum range of color variations).
No conclusion can be reached for LDN1415-IRS, and GM Cep, whose near-IR colors were obtained just once.
Similar JHK diagrams, for smaller sub-samples of eruptive stars, are also given by K\'{o}sp\'{a}l et al. (2011) and Kun et al. (2011b).

Summarizing, the majority of the newest objects present colors similar to those of classical ones, albeit extincted by the
expected (greater) amount.

\subsection{The optical two-colors diagram}

\begin{figure}
\includegraphics[angle=0,width=15cm] {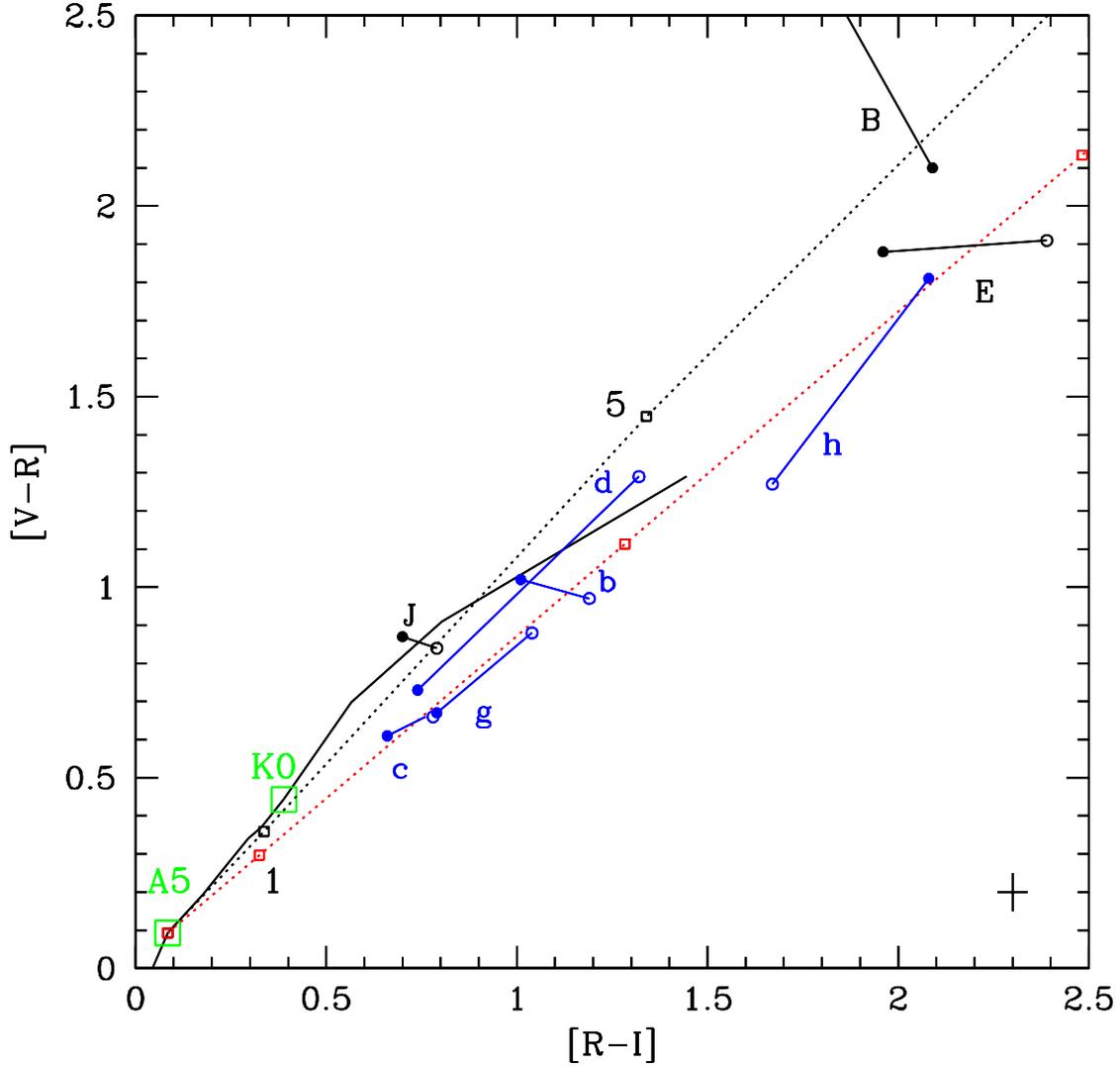}
   \caption{Visual two colors diagram of EXor in different epochs.
   The solid line marks the unreddened main sequence star of a given
   spectral type as defined by Kenyon \& Hartmann (1995) and corrected from Johnson
   to Cousins system (Fernie 1983); the relevant (see text)
   spectral types A5 and K0 are indicated as green open squares. The black and red
   dotted lines represent the reddening law by Rieke \& Lebofsky (1985)
   and Cardelli et al. (1989), respectively, where different values of A$_V$ are indicated by open
   squares. Sources are alphabetically coded as in Table 1. Solid (open) black
   circles identify the new targets during their outburst (quiescent) phase.
   The same is true for the classical EXor stars depicted in blue.
   Finally, a cross in the lower right corner indicates the typical error.
   \label{colcol_vis:fig}}
\end{figure}

It is worthwhile to repeat the above analysis by using the optical colors from V, R, and I, (when available) for which extinction and
scattering are both expected to play a major role (Figure~\ref{colcol_vis:fig}). In analogy with the near-IR diagram,
we indicate with blue (black) dots the points corresponding to the classical (newest) sources. For the reasons mentioned above, in the optical plot, the number of classical sources  exceeds that of the newest ones. Two of these latter (SVS 13 and V1647 Ori, labelled as B and E, respectively)) present color variations definitely not compatible with the predictions based on extinction, and the fluctuations exhibited by the classical cannot be accounted for with the A$_V$ values derivable from the near-IR colors.

Summarizing, the color analysis confirms the well established fact that phenomena different from extinction have a significant role in determining EXor's color variations. These latter, when derived by a single plot (VRI or JHK) could mimic an extinction variation which, however, is ruled out by the combined examination of both two-colors plots.

\subsection{Accretion parameters}

The comparison between classical and newest EXor can also benefit from the analysis of the mass accretion rate.
This parameter can be
derived by applying empirical relationships between the line luminosity and accretion luminosity (Antoniucci et al. 2011; Muzerolle et al. 1998).
This latter can be translated into a mass accretion rate (Gullbring et al. 1998) under some assumptions on the stellar parameters and the inner disk radius. This method relies on intrinsic line fluxes, therefore the lack of an accurate estimate of the extinction allows to
derive only approximated values for the accretion rate. The values available in the literature are collected in Table~\ref{Mdot:tab}.
Therein, the methods used for deriving $\dot{M}_{acc}$ are mentioned, giving also reference to the original papers where the adopted assumptions are explained. In the same Table, the status of the source when the accretion rate was derived is coded as high (H), intermediate (I), or low (L).

Classical sources show $\dot{M}_{acc}$ values in the interval 0.6$\cdot$10$^{-8}$ - 2.4$\cdot$10$^{-6}$ M$_{\sun}$ yr$^{-1}$, whereas the newest ones 1.6$\cdot$10$^{-8}$ - 1.8$\cdot$10$^{-6}$ M$_{\sun}$ yr$^{-1}$.
Hence, among classes of objects having similar masses (FUor, EXor, T Tauri), EXor's accretion rates appear comparable to T Tauri ones (typically 10$^{-7}$; Hartigan et al. 1995), but definitely lower than the typical rates of FUor objects (10$^{-5}$-10$^{-4}$; e.g. Hartmann \& Kenyon 1985).
Thus, no substantial difference is recognizable between the two classes, which appear once again very similar.

Data in Table~\ref{Mdot:tab}, combined with what we know about the recurrence of EXor events, allow us to infer on the total mass the star is capable to accrete through these repetitive events. By assuming, quite conservatively, ({\it i}) a peak value of the accretion rate of 2$\cdot$10$^{-6}$ M$_{\sun}$ yr$^{-1}$; ({\it ii}) that passing from outburst to quiescence is equivalent to maintain a constant accretion equal to 50\% of the peak value; ({\it iii}) that the outburst duration is 1/20 of the
quiescent stage; ({\it iv}) and finally, that the total evolutionary time of the EXor phase is $\leq$ 10$^{6}$ yrs, then the total mass accreted during such phase is M = 2$\cdot$10$^{-6}$ M$_{\sun}$ yr$^{-1}$ $\times$ 0.5 $\times$ 0.05 $\times$ ($<$10$^{6}$)yrs $\leq$ 0.05 M$_{\sun}$, at maximum.
Since Muzerolle et al. (1998) and Calvet et al. (2000) have already demonstrated that
the steady disk accretion rate cannot be responsible for a significant fraction of the final stellar mass, the FUor mechanism remains the sole alternative to accrete a substantial mass via disk accretion. However, at the largest possible rate of 10$^{-4}$ M$_{\sun}$ yr$^{-1}$, 10 or 100 FUor bursts, each lasting 10$^3$ or 10$^2$ yrs, respectively, are required. These prescriptions appear fairly extreme and they confirm that the final stellar mass is not quantitatively influenced by disk accretion during PMS.

\subsection{The outburst Spectral Energy Distribution}

Modeling the EXor SED's is a subject afforded and discussed in a number of dedicated papers
dealing with both individual sources (see references in Table~\ref{sample:tab}) and the entire class of objects (mainly classical, see eg. Herbig 2008). Typically, these models refer to peculiar phases (outbursts or quiescence)
or tend to interpret the overall observed phenomenology. Here we aim at comparing the SED corresponding to high and low states assumed concident with, or close to, the outburst and quiescent phase, respectively. Our aim is twofold:
{\it (i)} to understand if different outburst events can be taken back to a same observational phenomenology;
{\it (ii)} if so, to investigate how the derivable parameters can be used to infer
on a possible common mechanism. This scope looks quite ambitious, given
the limits imposed by the scanty observational material available so far, but a first effort in that direction
could be helpful for stimulating further investigations.

Aiming at reducing the number of approximations and assumptions, we have compared the SED's only at
those wavelengths where higher and lower state are both determined, so as to avoid any
interpolation at wavelengths where only one SED is given. Such an approach forced us to use less
photometric determinations than available, but assure us a more secure definition of the SED difference.
The fluxes we used for classical and newest EXor are given in Table~\ref{fluxes:tab}. These values are
based on minor assumptions: {\it (i)} the photometric bands have been {\it standardized}, in the sense that they
have been associated with a single wavelength, given in the Table notes, although the original fluxes were not
necessarily obtained with the same bandpass; {\it (ii)} the photometric errors adopted in the fitting procedure
are those given in the individual papers, otherwise, an uncertainty of 10\% is assumed.

We started our analysis by checking whether the SED variation between high and low state can be described through a simple change of the amount of extinction. In this scenario an intervening mass, which normally obscures the emitting system during quiescence, should be recurrenly removed during the outburst events. If this were true, the SED's ratio should resemble an extinction curve. Adopting the extinction law by Rieke \& Lebofsky (1985), we would obtain:

\begin{equation}
	\frac{SED^{out}(\lambda)}{SED^{quies}(\lambda)} = 10^{-0.4 \Delta A_{\lambda}}
\end{equation}

where: $\Delta A_{\lambda} = A_{\lambda}^{outburst}- A_{\lambda}^{quiescence}$, with $A_{\lambda} = A_V \left(\frac{5500}{\lambda}\right)^{1.75}$.

However, we find that it is impossible to fit the photometric observations with the previous function for any sources of our sample. This confirms the result of section 3.1, i.e. that the observed photometric variations must be ascribed to an intrinsic phenomenology more than to a simple extinction change.

Hence, we explored the hypothesis that an additional thermal component appears during the outbursting phase,
still allowing for a possible extinction variation between the quiescent and outburst stages.
In more detail, for each source we fitted the difference of the two de-reddened SED's (each corrected by the corresponding A$_V$ at outburst or quiescence) with a pure blackbody function (BB):

\begin{equation}
\label{eq:fit_BB}
	\frac{SED^{out}(\lambda)}{10^{-0.4 (A_{\lambda}+\Delta A_{\lambda})}} - \frac{SED^{quies}(\lambda)}{10^{-0.4 A_{\lambda}}} = \lambda \cdot BB(\lambda, T, R)
\end{equation}

where each SED is given as $\lambda F_{\lambda}$, A$_{\lambda}$ is the extinction at quiescence, R is the radius of the emitting region (assumed to be a uniform disk), and T is its BB temperature.
The free parameters are T, R, A$_V$, and $\Delta A_V$. We checked {\it a posteriori} that the derived  A$_V$, and $\Delta A_V$
values are roughly comparable with those given in the literature (see Table~\ref{sample:tab}). This is not the case of two
sources: V1647 Ori, and PV Cep, whose nature has been already considered quite uncertain (see Sect.2).
We point out that the shape of the true emitting area (that we do not model) is likely very different from the assumed uniform disk: for example it could be the inner wall of the circumstellar disk; nevertheless the radius R, derived by the fit, still provides an indication  of the length scale of the emission region.

\begin{figure}
\includegraphics[angle=0,width=15cm] {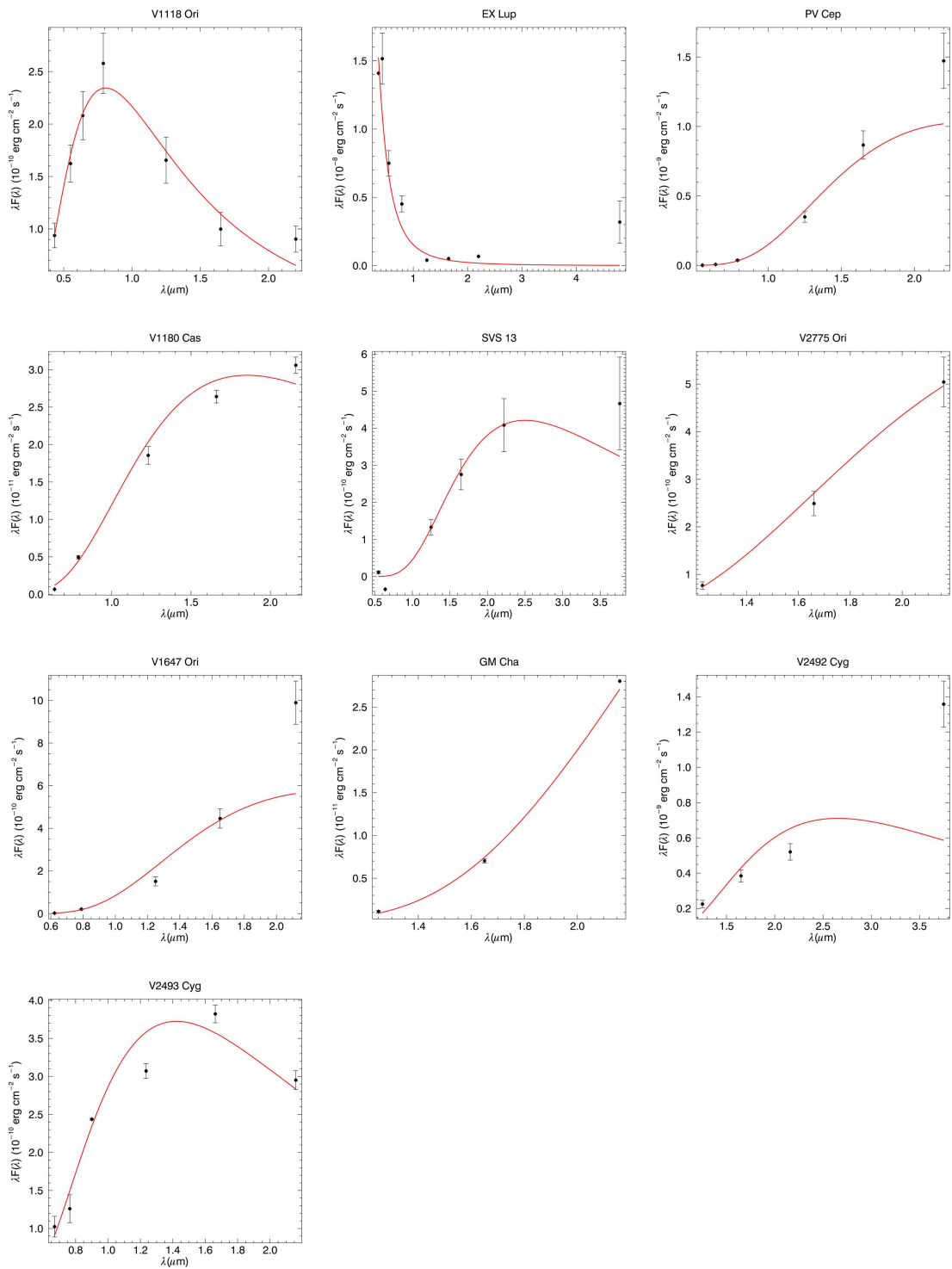}
   \caption{Best fits to the SED difference for both classical and newest EXor. Each curve represents a single temperature BB,
   whose value is given in Table~\ref{fit:tab}.
\label{fit:fig}}
\end{figure}

\begin{figure}
\includegraphics[angle=0,width=14cm] {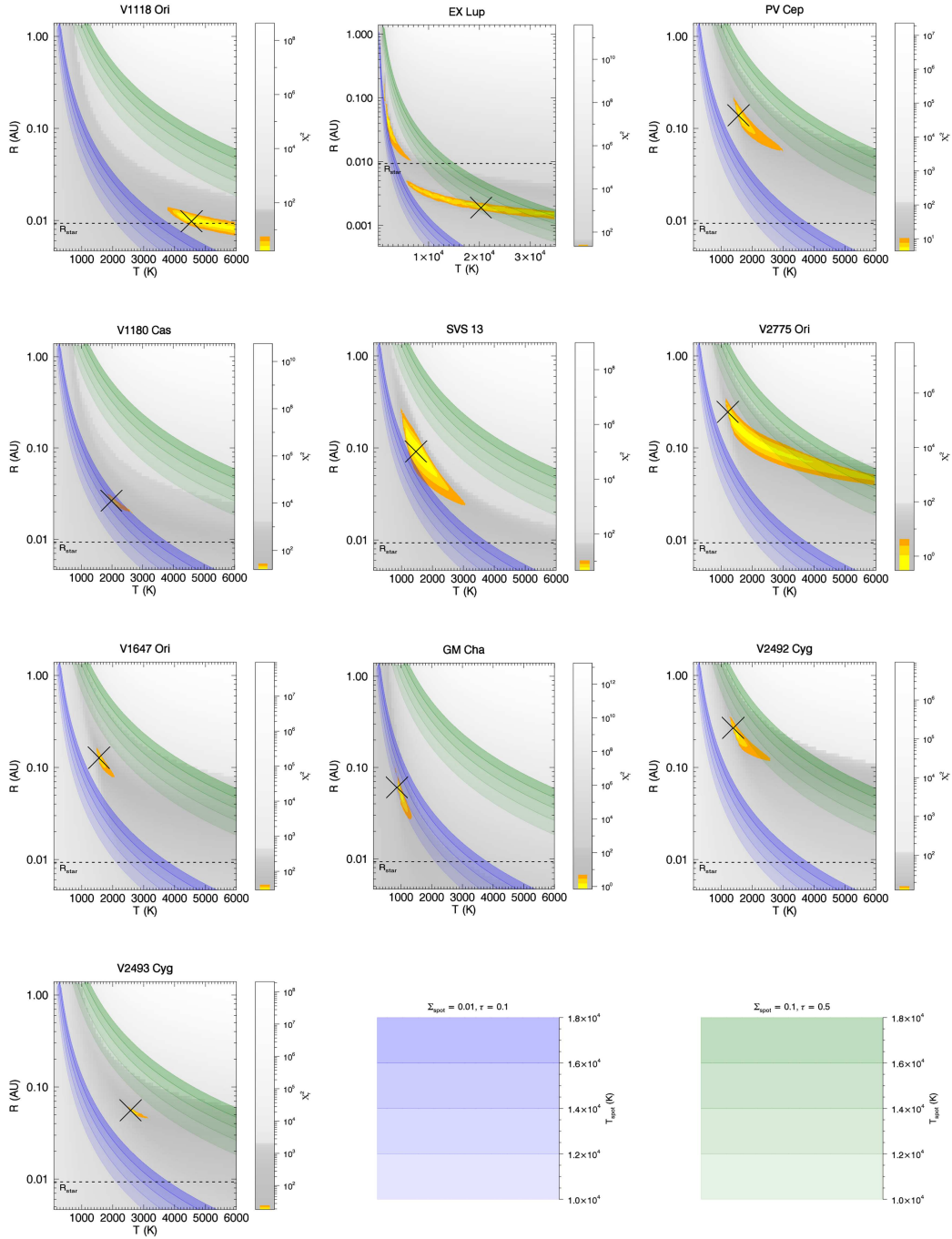}
    \caption{$\chi^2$ maps (grey-scale) projected on the (T, R) parameters plane with superposed 1$\sigma$ to 3$\sigma$ confidence contours (in yellow and orange, respectively). The dashed horizontal line indicates the stellar radius (assumed 2 R$_{\odot}$), while the shaded stripes show the parameters region allowed by the spot model and the energy conservation in a range of spot temperatures T$_{spot}$ (10000 K to 18000 K) for two values (10$^{-3}$ and 5~10$^{-2}$ in blue and green, respectively) of the $\tau\cdot\Sigma_{spot}$ product (see text). The color bars
    relative to these two values, for different spot temperatures, are given in the two bottom panels. The black crosses indicate the pair (R, T) corresponding to the absolute minimum of $\chi^2$.
\label{maps:fig}}
\end{figure}

Our results are depicted in Figure~\ref{fit:fig} and provide the indication that the appearance of a thermal component is a good representation of the observed phenomenum. Almost all the sources present a definite single temperature component spanning between 1000 K and 4500 K. Two exception can be noted: {\it (i)} EX Lup shows a higher temperature ($\sim$20000 K) BB coming from a very compact region (fraction of stellar radius): this result is not consistent with the value of 6500 K by Juh\'{a}sz et al. (2011) and this discrepancy will be discussed below; {\it (ii)} V2492, whose excess emission is barely fitted with a single BB (se also K\'{o}sp\'{a}l  et al. 2011). The temperature we derive for V1647 Ori is not conflicting with those by McGehee et al. (2004) and Walter et al. (2004), since here we are dealing with an excess temperature component instead of the
T value at outburst. Minor corrections to the pure BB could be applied by accounting for the excess continuum emission observed in T Tauri through the IYJ bands (Fischer et al. 2011), but the existence of multiple contributions at substantially different temperatures seems to be ruled out. The output parameters of our fits are given in Table~\ref{fit:tab}. As can be seen in our $\chi^2$ maps
(see Figure~\ref{maps:fig} and discussion below), there exists a range of temperatures T and radii R providing fits with a comparable $\chi^2$ value. These ranges are given in Table~\ref{fit:tab} for each source. Noticeably, such R values are smaller than the typical radius of the inner portion of accretion disks (Muzerolle et al. 2003), but this is expected, since we are assuming that the emission is ideally coming from a face-on disk centered on the star.
Moreover, in Table~\ref{fit:tab} we see that the extinction variations $\Delta$A$_V$ always present small negative values, the average value being -3.1 $\pm$ 1.8, which means that the outburst tends not only to heat, but also to remove and sublimate some dust (see below).
The large spread of temperature values likely reflects the large variety of morphological and physical conditions (inner disk size, opacity, density) more than being suggestive of a random sampling of a dynamical process at  different epochs (e.g. the heating or cooling of a circumstellar region). To test this hypothesis, we applied our fit method also to SED's relative to intermediate epochs of the two sources (V1118 Ori and V1647 Ori) whose outburst periods are well covered by several
optical/near-IR observations. If we looked at a progressive phenomenum, we should obtain temperatures increasing or decreasing with fluxes.
Conversely, we obtain the same value of the temperature by fitting consecutive SED's which are not substantially
far in time from the outburst. However, as time goes on, a single BB appears more and more inadequate to account for the data. This likely means
that the existence of a single BB is well representative of the outburst, while a stratification of temperatures is more suitable to describe the intermediate phase toward the quiescence.

We tried to interpret this observational scenario by considering that the EXor accretion event leads to the onset of a hot spot on the stellar surface. The spopt both warms the circumstellar region, whose thermal emission we are observing, and also contributes to dust sublimation. We rely on energy conservation by assuming that the energy irradiated by the spot equals the sum of that emitted by the surrounding region and that spent for dust sublimation. In practice, adopting an efficiency equal to unity, the spot power (i.e. luminosity) integrated for the spot duration should be equal to the total energy emitted during all the outburst by the inner disk (heated by the spot itself) plus the energy needed for sublimation.


An estimate of this latter can be obtained for a typical T Tauri star (M=0.5 M$_{\sun}$, R=2 R$_{\sun}$, T=4000K). The sudden increase of $\dot{M}_{acc}$ from 0.02 to 6 10$^{-7}$ (assuming as typical the values of EX Lup - see Table~\ref{Mdot:tab}) causes the accretion luminosity to increase from 0.05 to 1.58 L$_{\sun}$. The dust inner disk which is located at about R$_1$=17 R$_{\sun}$ in quiescence, increases up to R$_2$=30 R$_{\sun}$ during the outburst (Muzerolle et al. 2003): obviously, the burst heats the inner zone and, consequently, moves the dusty disk away.
We can find an upper limit of the energy spent for dust sublimation by assuming: {\it (i)} the dust is contained in the entire volume between R$_1$ and R$_2$; {\it (ii)} such dust evaporates completely. More precisely the sublimating dust is located within a cylinder having a radial width R$_2$-R$_1$ and scale height (H) related to the vertical density variation from the disk plane. Following Muzerolle et al. (2004), such volume (V) and the internal gas mass (m) are given by:

\begin{equation}
\label{eq:vol_mass}
	V = 4 \pi (R_2^2 - R_1^2) \cdot H{~~~~~~~~~~~~~~~~~~~}m = \rho \cdot V
\end{equation}

where: H=c$_s$/$\omega$ is the height scale, c$_s$ is the sound speed (given by c$_s^2$= KT/$\mu$m$_H$), $\omega$ is the keplerian velocity $\omega^2$ = GM/R$_1$ , and $\rho$ is the volume density of the inner disk, that can be obtained in terms of the surface density $\sigma$ and the height scale H, as $\rho$ = $\sigma$/2$\pi$H. With the numerical values for
T, R$_1$, and M given above, we obtain m = 5$\cdot$10$^{27}$ g, which represents a large upper limit since neither the density decreasing from R$_1$ and R$_2$, nor its vertical decreasing have been accounted for.
Now, assuming a solar chemistry, half of the silicon locked onto olivine grains (olivine is the most stable silicate, see
Gail \& Sedlmayr 1999), and adopting the compilation of Sharp \& Huebner (1990, their Table 2) to obtain the energy absorbed
for the sublimation, we find that the total energy needed to sublimate the dust present in the inner disk is  6.4$\cdot$10$^{35}$ erg. Such amount of energy is released away in a very short time (minutes) and only a tiny fraction hits the inner disk, therefore it can be neglected in the energy budget.

The evaluation of the spot emission is done following the model by Calvet \& Gullbring (1998) which predicts spot temperatures (T$_{spot}$) between 6000 and 12000 K for different combination of stellar mass and radius and for a given accretion rate in
the range of 10$^{-9}$-10$^{-7}$ M$_{\sun}$ yr$^{-1}$. The spot temperature roughly scales as $(\dot{M}_{acc})^{1/4}$, hence the EXor events, whose mass accretion rates span between 2$\cdot$10$^{-7}$ and 2$\cdot$10$^{-6}$ M$_{\sun}$ yr$^{-1}$ (see Table~\ref{Mdot:tab}), should create spots with temperatures between 10000 and 18000 K, for typical values of stellar mass and radius (0.5 M$_{\sun}$, 2 R$_{\sun}$, respectively). Such spot temperatures give a barely detectable emission in the optical near-IR bands, which is not easily recognizable in the SED's difference.


The power emitted by the spot linearly depends on two factors: the time interval $\tau$ during which the spot temperature remains at its maximum value (10000-18000K), expressed as a fraction of the outburst duration, and the spot surface $\Sigma_{spot}$,  expressed as a fraction of the stellar surface. Given such linearity, we treated the product $\tau \cdot \Sigma_{spot}$ as a single variable. For the given range of spot temperatures (from 10000 to 18000 K), two extreme values of the product $\tau \cdot \Sigma_{spot}$ are indicated in blue and green in Figure~\ref{maps:fig}, corresponding to 0.001 ($\tau$ = 0.1 $\times$ $\Sigma_{spot}$ = 0.01) and 0.05 ($\tau$ = 0.5 $\times$ $\Sigma_{spot}$ = 0.1), respectively. The considered upper value for $\tau$ is compatible with the typical light curve (see e.g. V1118 Ori - Audard  et al 2010) where the maximum flux level lasts about half of the total outburst duration. The lower value for $\Sigma_{spot}$ are compatible with the sizes predicted by Calvet \& Gullbring (1998).

In Figure~\ref{maps:fig}, the area of intersection between the
$\chi^2$ minima and model predictions represent the T and R ranges for which the spot model can account for the observations.
The derived ranges of T and R are depicted in Figure~\ref{maps:fig} as confidence contours (in yellow-orange) superposed on the $\chi^2$ maps and are listed in Table~\ref{fit:tab}.

Summarizing, for 8 out of the 10 investigated objects we find a BB fit in very good agreement with model predictions. Two extreme cases are EX Lup, which can be interpreted as a direct detection of hot spot emission, and PV Cep, whose observations are compatible with a marginal value of the emitting region radius (R = 0.2 AU).

Hence, the increase of luminosity during the EXor event appears compatible with emission from an inner disk heated by a hot spot. It is worthwhile noting that Calvet \& Gullbring (1998) model predicts an additional contribution to the heating (of about 30\%) due to the column above the accretion shock (i.e. the pre-shock region): here, we assume it compensates possible efficiency losses currently ignored. This quantitative analysis of EXor events provides further support to the magnetospheric accretion infall model.

\section{Comparison with CTTSs}

\begin{figure}
\includegraphics[angle=0,width=15cm] {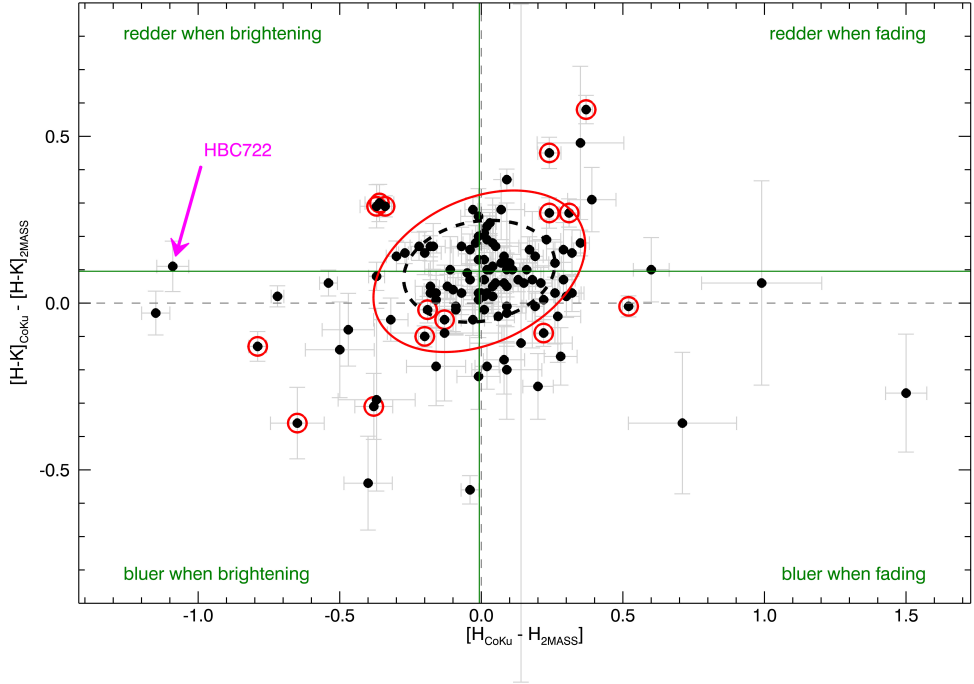}
   \caption{Distribution of the color [H-K] variations
   vs. the H magnitude variations between early near-IR observations (Cohen \& Kuhi 1979) and 2MASS photometry (Cutri et al.
   2003) of a sample of low mass YSO's (mainly CTTSs, see Sect.2). All the sources are depicted
   with their uncertainty. Among them, circled dots identify sources that (simultaneously) present both magnitude and
   color variation at a S/N level $>$ 5 and $>$ 2.5, respectively. Different quadrants are associated with different
   modalities of variation, shortly indicated in each corner (see also text). The continuous (dashed) ellipse represents 1$\sigma$ of the distribution of the plain (circled) dots. The nominal cartesian vertical and
   horizontal axes are indicated as dashed lines, but the real ones (solid axes) have been determined using the data points
   barycenter (see text).
   \label{variables:fig}}
\end{figure}

The second aim of our work is to search for similarities (or differences) between the photometric behaviour of the
known EXor and those of randomly variable CTTSs. To perform a statistically significant comparison, we have searched
the literature data for a large sample of sources for which there exist two photometric measurements, taken almost simultaneously in two different epochs. We selected the sample investigated by Cohen \& Kuhi (1979) (hereafter CoKu), who
observed through near-IR photometry and optical spectroscopy a large number of low-mass PMS sources belonging to different clouds.
Twenty years later, during the 2MASS survey, practically all these sources were re-observed in the same bands.
Data from both these surveys offer the chance of analyzing the modalities of the occurred
variations (H, K bands) for 134 objects on a time-scale that, although not adequate to infer on the rising and
declining duration of possible outburst events, is still suitable for a statistical
investigation. Aiming at recognizing objects characterized by irregular variability,
typically associated with eruptive events, we have excluded the sources
potentially associated with periodic variations. Following the robust results by Herbst et al.(1994),
we have excluded WTTSs (characterized by W$_{H\alpha} <$ 10\AA) mainly exhibiting cool spots, and we have considered only CTTSs (SpT. later than K0 and W$_{H\alpha} >$ 10\AA). Our final sample is composed by 120 objects, whose color variation as a function of the H magnitude variation is depicted in Figure~\ref{variables:fig}.\

We note that CoKu and 2MASS photometry were obtained with similar H-band
filters ($\lambda_{eff}$= 1.65 and 1.66 $\mu$m for CoKu and 2MASS, respectively), but with different
K-band filters ($\lambda_{eff}$= 2.31 and 2.16 $\mu$m for CoKu and 2MASS, respectively). This does not
affect the difference between the two H magnitudes, but introduces systematic effects in the
[H-K] color differences. We did not attempt to perform a color correction, which would be based on {\it ad hoc} assumptions;
instead, we take as (0,0) point the barycenter of the data distribution. Indeed, by looking
at the Figure~\ref{variables:fig}, we can see that the abscissa origin {\it indicated} by the data-point distribution
(vertical green line) is practically identical to the nominal origin (dashed vertical line), while the same is not true for
the ordinate axis, where a difference of about 0.1 mag is recognizable.\\
We identify with black dots all the 120 sources and with red circled those presenting variations above 5$\sigma$ in [$\Delta$H] and 2.5 $\sigma$ in [$\Delta$(H-K)].
Correspondingly, black and red ellipses indicate the 1$\sigma$ distribution of the two groups. From the inspection of Figure~\ref{variables:fig} we can draw, although qualitatively, some noticeable statements: while the black ellipse does not show any peculiar orientation, the red one has its major axis definitely oriented toward the 1st and 3rd quadrants, that is those populated by sources that become bluer when brightening and redder when fading. The same behaviour is exhibited by EXor events (Lorenzetti et al. 2009 and references therein): therefore, in the framework of the PMS evolution, disk accretion phenomena, like those responsible for EXor bursts, seem to have a role also in accounting for the more frequent and less strong photometric variations typical of all CTTSs.

According to the above picture, EXor could represent rarer cases of a very common and widespread phenomenology displayed by all the CTTSs. This view is also supported by the accretion model recently presented by D'Angelo \& Spruit (2010), who suggest the existence of a common mechanism that regulates both CTTS and EXor accretion events. Indeed, their predictions are able to explain both the strongest (EXor) and the weakest (CTTS) episodes of accretion, which would be determined by the interplay between the amount of gas that piles up in the inner region of the disk and the radius at which the keplerian frequency in the disk equals the star rotational frequency. The cyclic
accretion solutions continue to exist indefinitely, also when the accretion rate decreases.

As a second output of our analysis, here we provide the identification of the sources that in Figure~\ref{variables:fig}
display the largest photometric (and color) variations since, for that reason, they could be targeted in forthcoming
observational programs. They all presented fluctuations $\Delta$H $\simeq$ 1 mag (or more) as V2493 Cyg (see next Sect.), and, noticeably,
three of them, LkH$\alpha$154, LkH$\alpha$175 and LkH$\alpha$186 belong to the same cloud NGC7000/IC5070,
while other three (NX Mon, LR Mon, and OY Mon) are located in NGC 2264. The circumstance that the most active CTTS
cluster within only two well defined groups could be not fortuitous, but signals a larger frequency of accreting bursts in objects of the younger regions. In particular, NGC2264 (NGC7000) variables lie within 16 (65) arcmin, which corresponds to 0.06 (0.17) pc; the mutual closeness of these objects supports their coevality and, in turn, the possibility that they are suitable EXor candidate, if such a phase were typical of the early evolution of any CTTS.

Several (complete) samples of T Tauri stars have been recently monitored at IR frequencies (e.g.YSOVAR, Morales-Calder\'{o}n et al. 2011; PAIRTEL, Bloom et al. 2006; the surveys by Carpenter et al. 2001; Giannini et al. 2009; Alves de Oliveira et al. 2008) all aiming at establishing some statistics on the most relevant types of variability: i.e. rotational modulation by cool or hot spots, obscuration by dust, eruptive accretion. Strongly eruptive EXor are rare, but the identification of a couple of such objects (V2493 Cyg and V2492 Cyg) was provided by the mentioned surveys, which therefore could be effectively mined to determine plausible candidates.

\section{The case of V2493 Cyg}

Almost all the known EXor present frequent and smaller outbursts between two major events (see e.g. the 
behaviour of prototype EX Lup between the 1955 and 2008 outbursts - Aspin et al. 2010). Moreover, some EXor events might have occurred during the periods of weaker activity (of V2493 Cyg,
but also of any other CTTS), but they might have been undetected because of the lack of a systematic monitoring. This latter consideration suggests that EXor events could be much more frequent than suggested by the few  cases accidentally observed. An illustrative case
is represented by the source indicated with an arrow in Figure~\ref{variables:fig}. This is V2493 Cyg ({\it alias} LkH$\alpha$188G4), classified as CTTS
until the summer 2010, and now recognized as an EXor object (K\'{o}sp\'{a}l et al. 2011), or, possibly, as a FUor candidate
mainly because of the presence of absorption lines in its optical and near-IR spectrum (Semkov \& Peneva 2010, Miller et al 2011).
The fact that V2493 Cyg was a CTTS undergoing smaller photometric variations (typically 1-2 mag) was obviously known since the
2MASS epoch.

What happened to V2493 Cyg (and to other observed CTTSs in outburst) could lead to the speculative
conclusion that all CTTSs undergo an EXor phase. This latter could take place in the early CTTS lifetime, when the accretion rate is expectedly higher, or it could randomly appear during the subsequent evolution: the model of D'Angelo \& Spruit tends to support the former hypothesis. Surely, given the scarcity of the observed cases, the EXor phase should be a short-lasting period of the PMS evolution.

Given the considerable interest provoked by its recent outburst discovery, we included V2493 Cyg among the targets of our 
EXor variables monitoring program. Our observations partially cover a period of about one year after the 2010 August outburst.
Optical photometric (and very few polarimetric) observations were obtained
with two nearly-identical photometers-polarimeters of the
Astronomical Institute of St.Petersburg University, mounted on the
70cm AZT-8 telescope of the Crimean Observatory (Ukraine) and 40cm
LX-200 telescope of St.Petersburg University, respectively.
Near-IR photometry and spectroscopy were obtained at the 1.1m AZT-24 telescope located at
Campo Imperatore (L'Aquila - Italy). The exploited instrumentation along with the procedures
for data acquisition and reduction are extensively described in Lorenzetti et al. (2006).
Noticeably, the intra-day or day time-scale variations are
not relevant for the discussion, so that observations in different bands
not strictly simultaneous have been associated to the same date (within a maximum distance of 1 day).
Photometric data are given in Table~\ref{HBC722_data:tab}: they do not stem from a systematic monitoring.
Our earliest observation is the first one of V2493 Cyg in outburst (Leoni et al. 2010), then a slow fading phase
started, even if the quiescence pre-outburst state (corresponding to about 12 mag in H band) has never been reached.

\begin{figure}
\includegraphics[angle=0,width=15cm] {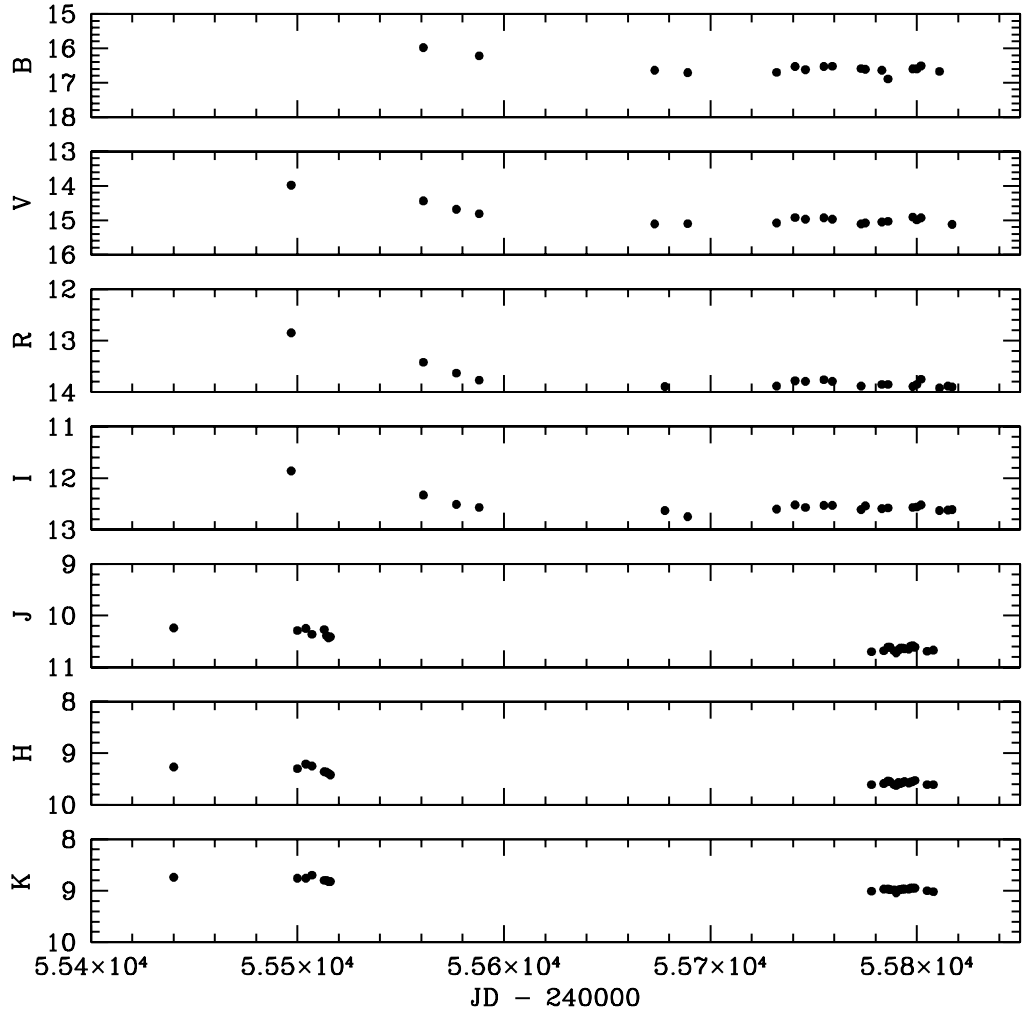}
   \caption{V2493 Cyg optical and near-IR light curves vs. Julian Date. The errors
   of data points are comparable to or less than 0.04 mag.
\label{lightcurve:fig}}
\end{figure}

Optical and near-IR data are presented in  Table~\ref{HBC722_data:tab} and depicted as light-curves in Figure~\ref{lightcurve:fig}. These latter indicate that the source, after an early declining more pronounced in optical (\lapprox~1 mag) than in near-IR bands ($\approx$ 0.5 mag), is in a steady state since about ten months. Some near-IR (I to K band) spectra of V2493 Cyg have been also taken during the monitoring period: given the
limits imposed by our instrumentation, they appear quite noisy, however the systematic trend to become redder and redder while fading is recognizable. Moreover, two spectral features present in more than one spectrum can be identified (at a significant S/N level): the Pa$\beta$ HI at 1.282$\mu$m and the CO band (2$\rightarrow$0) around 2.35$\mu$m; remarkably, they both appear always in absorption (see also Miller et al 2011). The persistence of a high flux level after the outburst and the presence of absorption features are defining properties of FUor objects more than of EXor, which tend to go back to their quiescent state within a short time
(about 1 year) and usually present emission line spectra. Hence the possibility that V2493 Cyg underwent a FUor event cannot be ruled out, although
the outburst mass accretion rate (see Table~\ref{Mdot:tab}) is significantly lower than that found in FUor systems. Finally, we acquired several I band polarimetric measurements of this source, finding that it is remarkably stable at the level of 3-4\%; the derived position angle has always the same value of 158$^{\circ}$ $\pm$ 2$^{\circ}$, which should correspond to the disk orientation. The rather low polarization value is typical of a system viewed pole-on, or, in any case, of a system whose phenomenology is not disk-dominated, as in FU Ori events. Indeed, in this case, a small fraction of the light from the star is attenuated and the scattered/polarized light is typically low. In conclusion, to ascertain its real nature, V2493 Cyg certainly deserves a further monitoring during, at least, the next two years.

\section{Concluding remarks}

We have performed a systematical analysis based on literature data of both classical and newest identified EXor,
mainly aiming at revealing similarities or differences between the two categories. In particular, we concentrated on
optical/near-IR photometric data relative to two specific phases: outburst or quiescence. By comparing the SED's
of both phases we infer on a possible mechanism responsible for the outburst itself. Based on our analysis, the
following results can be summarized:

\begin{itemize}
\item[1.] In the [J-H],[H-K] two-color diagram the newest sources appear indistinguishable with respect to the classical ones, suggesting that no intrinsic differences exists between the two categories. Only minor differences can be noted, due both to the embedness of the newest sources located in more extincted regions, and to a more complete observational sampling.
\item[2.] Optical [V-R],[R-I] and near-IR colors confirm that phenomena different from extinction may be significant in explaining the color variation associable with EXor outbursts (both classical and new).
\item[3.] The mass accretion rate ($\dot{M}_{acc}$) exhibited by classical EXor span a slightly wider range with respect to the newest objects (0.006-2.4 vs. 0.016-1.8 10$^{-6}$ M$_{\sun}$ yr$^{-1}$), but no substantial difference is recognizable between the two classes, which appear once again very similar. They both present $\dot{M}_{acc}$ values that are comparable with the classical T Tauri ones and never reach the rates typical of FUor objects (10$^{-5}$-10$^{-4}$). When outbursting, EXor objects tend to have accretion rates an order of magnitude greater than in quiescence.
\item[4.] We computed the difference between the SED at high and low state (close to outburst and quiescence, respectively) for both classical and newest sources. All these differences are well fitted with a single blackbody whose temperature spans in the range 1000-4500 K, while the radii of the emitting regions span between 0.01 and 0.1 AU.
\item[5.] By adopting the hot spot model by Calvet \& Gullbring (1998), the appearance of an additional blackbody during the outburst (up to few hundreds of days) can be interpreted in terms of heating of the inner parts of the accretion disk by a hot spot of 10000-18000 K having a size not exceeding 10\% of the stellar surface.
\item[7.] A comparison with a large sample of active T Tauri stars shows that the most significant variations (both in magnitude and  color) of these latter present the same modalities of EXor variations, although to a lesser extent. Hence, accretion phenomena have a dominating role also in the complex scenario of the CTTS variability.
\item[8.] This comparison leads us to speculate that EXor events might represent an infrequent manifestation of a rather common phenomenology displayed by all CTTSs. In this context, the rarity of EXor objects could be just due to the lack of a systematic monitoring.
\item[9.] In particular, one of the considered CTTSs, namely V2493 Cyg, underwent a strong accretion event in 2010 August, whose EXor or FUor nature is still controversial. Given the interest on this source, it was included in our list of monitoring program. Optical and near-IR observations (obtained in the period 2010-2011) have been presented. Basing on these preliminary data, its FUor nature cannot be ruled out, although a longer-lasting monitoring (at least a further couple of years) is needed for ascertaining its real nature.
\end{itemize}

\section{Acknowledgements}

We would like to thank the anonymous referee who significantly contributed to improve the present paper. We also thank
Mauro Centrone and Riccardo Leoni for their support to the Campo Imperatore observing runs.

{}

\begin{figure}
\includegraphics[angle=0,width=15cm] {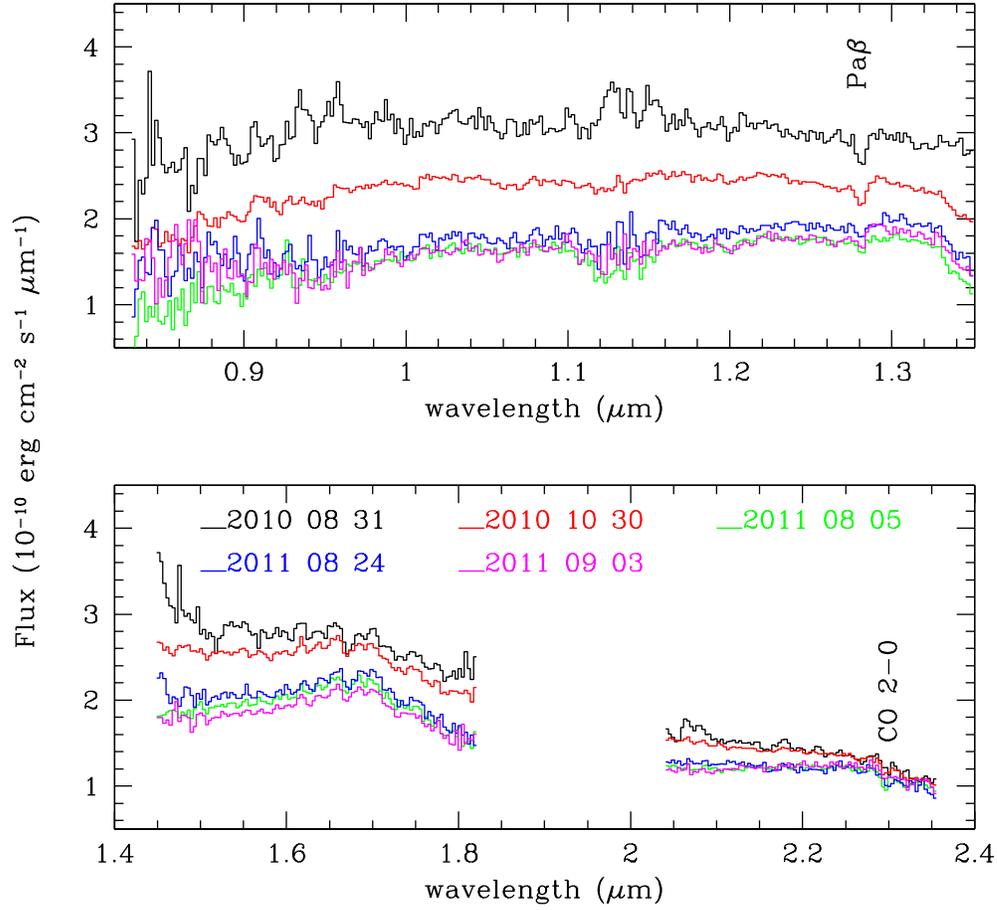}
   \caption{Near-IR spectrum of V2493 Cyg obtained on different dates. Detected features are identified.
\label{spectrum:fig}}
\end{figure}

\begin{deluxetable}{lcccccccc}
\tabletypesize{\footnotesize} \tablecaption{EXor (classical and
new detections) relevant observed parameters. \label{sample:tab}}
\tablewidth{0pt}
\tablehead{Source & Id.$^a$ &Distance&Location& A$_V$      & L$_{bol}$  & V$_{max}$ & V$_{min}$     & References \\
                  &         &  (pc)  &        & (mag)      &(L$_{\sun}$)& \multicolumn{2}{c}{(mag)} &             }
\startdata
UZ Tau E           & a &   140  & B19    &  1.49      & 1.7 $^b$   &   11.7    &  15.0         & 1,2        \\
VY Tau             & b &   140  & L1536  &  0.85      & 0.75$^b$   &    9.0    &  15.3         & 1,2,3      \\
DR Tau             & c &   140  & L1558  & 1.7-2.1    & 1.05-5.0   &   10.5    &  16.0         & 1,2,4,5    \\
V1118 Ori          & d &   414  & ONC    &   0-2      & 1.4-25.4   &   12.8    &  17.5         & 1,6,7      \\
NY Ori             & e &   414  & ONC    &   0.3      &   ...      &   14.5    &  17.5         & 1,6,8      \\
V1143 Ori          & f &   500  & L1640  &   ...      &   ...      &    13     &  19           & 1          \\
EX Lup             & g &   140  & Lup3   &    0       &   0.73     &    8.4    &  13.2         & 1,9,10,11  \\
PV Cep             & h &   325  & L1158  &  5-7       &   100      &   14.6    &  18.0         & 12,13,14,15,16\\
\hline
V1180 Cas          & A &   600  &L1340         &   4.3      &   0.07     &  15.7$^e$ &  $>$ 21  & 17   \\
SVS13              & B &   300  &NGC1333       &   6-15     &   ...      &  15.9$^e$ &  19.0    & 18,19,20\\
LDN1415-IRS        & C &   170  &LDN1415       &   ...      &   ...      & 14.74$^e$ & 18.37    & 21   \\
V2775 Ori          & D &   450  &L1641         &    18      & 1.9-22     & 11.77$^f$ & 16.4     & 22   \\
V1647 Ori          & E &   400  &L1630         &   9-19     & 5.2$^b$    & 14.40$^e$ & 20.27    & 23,24\\
                   &   &        &              &            & 2.8-7.7$^c$&           &          & 24   \\
                   &   &        &              &            & 44$^d$     &           &          & 25   \\
GM Cha             & F &   160  &Cha I         &    13      &   1.5      & 10.61$^g$ & 12.75    & 26   \\
OO Ser             & G &   311  &Ser NW        &    42      & 4.5-(26/36)& 11.4$^g$  & 16.1     & 27   \\
V2492 Cyg          & H &   600  &IC5070        &   6-12     &    20      & 14.7$^h$  & 18-19    & 28,29,30\\
V2493 Cyg          & I &   600  &NGC7000/IC5070&   3.4      & 2.7-12     &   13.62   & 17.0     & 4    \\
GM Cep             & J &   900  & Tr 37        &   2-4      &   30/40    & 12.43$^h$ & 14.58    & 31   \\
\enddata
\tablenotetext{a}{Individual sources are identified with the same alphabetical letter used in the colors plots}
\tablenotetext{b}{~~photospheric luminosity}  \tablenotetext{c}{~~low state accretion luminosity}
\tablenotetext{d}{~~outburst bolometric luminosity}
\tablenotetext{e}{~~I band} \tablenotetext{f}{~~J band}
\tablenotetext{g}{~~K band} \tablenotetext{h}{~~R band}

\tablecomments{References to the Table: (1) SIMBAD Astronomical
Database (http://simbad.u-strasbg.fr/simbad); (2) Herbig \& Bell 1988;
(3) Herbig 1990; (4) Cohen \& Kuhi 1979; (5) Kenyon
et al. 1994; (6) Menten et al. 2007; (7) Audard et al. 2010; (8) Breger, Gherz \&
Hackwell 1981; (9) Gras-Vel\'{a}zquez \& Ray 2005; (10) Herbig et
al. 2001; (11) Sipos et al. 2009; (12) Cohen et al. 1981; (13) Van Citters \& Smith 1989;
(14) Lorenzetti et al. 2011; (15) Straizys et al. 1992; (16) Kun et al. 2011a
(17) Kun et al. 2011b; (18) Takami et al. 2006; (19) Liseau
et al. 1992; (20) Eisl\"{o}ffel et al. 1991; (21) Stecklum et al.
2007; (22) Caratti o Garatti et al. 2011; (23) Brice\~{n}o et al.
2004; (24) Aspin et al. 2008; (25) Muzerolle et al. 2005;
(26) Persi et al. 2007; (27) K\'{o}sp\'{a}l et al. 2007;
(28) Covey et al. 2011; (29) Aspin 2011; (30) K\'{o}sp\'{a}l et al. 2011;
(31) Sicilia-Aguilar et al. 2008;}

\end{deluxetable}

\begin{deluxetable}{lcccc}
\tabletypesize{\footnotesize} \tablecaption{EXor mass accretion rates in different phases. \label{Mdot:tab}}
\tablewidth{0pt}
\tablehead{ Source & Status$^a$ &  $\dot{M}_{acc}$ (10$^{-7}$ M$_{\sun}$ yr$^{-1}$) & Ref & Method  } \startdata
UZ Tau E         & I & 1-3                 & 1   & Pa$\beta$              \\
DR Tau           & I & 4-9                 & 1   & Pa$\beta$, Br$\gamma$  \\
V1118 Ori        & H & 10 $\pm$ 5          & 2   & SED modeling  \\
                 & L & 2.5                 & 2   & SED modeling  \\
EX Lup           & H & 2 $\pm$ 0.5         & 3   & Br$\gamma$    \\
                 & L & 0.06 $\pm$ 0.03     & 3   & Pa$\beta$     \\
PV Cep           & H & 24                  & 1   & Pa$\beta$, Br$\gamma$  \\
                 & I & 19                  & 1   & Pa$\beta$, Br$\gamma$  \\
                 &   &                     &     &               \\

\hline
V1180 Cas        & H & $>$ 1.6             & 4   &  CaII $\lambda$8542           \\
                 & L & $>$ 0.16            & 4   &  estimated by Lum. scaling    \\
V1647 Ori        & H & 10 $\pm$ 5          & 5   &  Near-IR lines,               \\
                 &H/I& 5.0                 & 6   &  Near-IR lines,               \\
                 & L & 0.55                & 6   &  Near-IR lines,               \\
V2492 Cyg        & H & 6.4-18              & 7   &  Br$\gamma$                   \\
                 & H & 2.5                 & 8   &  Near-IR lines                \\
V2493 Cyg        & H & 10                  & 9   &  estimated by Lum. scaling    \\
                 & L & 0.5                 & 10  &  estimated by the given parameters \\
\enddata
\tablenotetext{a}{H = high; I = intermediate; L = low}
\tablecomments{References to the Table: Lorenzetti et al. 2009; (2) Audard et al. 2010; (3) Aspin et al. 2010; (4) Kun et al. 2011; (5) Muzerolle et al. 2005; (6) Acosta-Pulido et al. (2007); (7) Aspin 2011; (8) Covey et al. 2011; (9) K\'{o}sp\'{a}l et al. 2011; (10) Cohen \& Kuhi 1979.}

\end{deluxetable}

\begin{deluxetable}{lccccccccc}
\tabletypesize{\footnotesize} \tablecaption{Fluxes in high and low state. \label{fluxes:tab}}
\tablewidth{0pt}
\tablehead{ Source & O/Q$^a$ &  \multicolumn{7}{c}{$\lambda\cdot$F$_{\lambda}$  (erg s$^{-1}$ cm$^{-2}$)} & Ref. \\
                   & &  B$^b$  & V       & R       & I       & J       & H       & K            &   }
\startdata
V1118 Ori          &+&9.09(-11)&1.47(-10)&1.99(-10)&2.62(-10)&2.19(-10)&1.72(-10)&1.29(-10)     &  1,2    \\
                   &-&1.35(-12)&2.15(-12)&4.87(-12)&1.10(-11)&3.99(-11)&5.74(-11)&3.42(-11)     &  1,2    \\
EX Lup$^c$         &+&2.22(-9) &2.28(-10)&    ---  &2.86(-9) &9.59(-10)&9.61(-10)&9.13(-10)     &  3      \\
                   &-&5.63(-11)&1.11(-10)&    ---  &3.43(-10)&4.84(-10)&4.74(-10)&2.82(-10)     &  4      \\
PV Cep             &+&   ---   &2.66(-12)&9.71(-12)&4.42(-11)&3.69(-10)&1.00(-9) &2.04(-9)      &  5      \\
                   &-&   ---   &1.14(-12)&2.53(-12)&7.90(-12)&3.01(-11)&1.48(-10)&5.63(-10)     &  5      \\
\hline
V1180 Cas          &+&   ---   &   ---   &3.52(-12)&6.33(-12)&2.06(-11)&3.04(-11)&3.97(-11)     &  6     \\
                   &-&   ---   &   ---   &1.51(-13)&2.12(-13)&9.81(-13)&2.48(-12)&6.58(-12)     &  6,7   \\
SVS 13$^d$         &+&   ---   &1.71(-13)&8.15(-13)&   ---   &5.58(-11)&1.88(-10)&4.95(-10)     &  8,9   \\
                   &-&   ---   &6.31(-15)&1.09(-13)&   ---   &1.11(-11)&4.60(-11)&1.92(-10)     &  8,9   \\
V2775 Ori          &+&   ---   &   ---   &   ---   &   ---   &7.58(-11)&2.53(-10)&5.18(-10)     &  10    \\
                   &-&   ---   &   ---   &   ---   &   ---   &1.07(-12)&7.15(-12)&1.83(-11)     &  10    \\
V1647 Ori          &+&   ---   &   ---   &1.92(-12)&1.68(-11)&1.50(-10)&4.64(-10)&1.06(-9)      &11,12,13\\
                   &-&   ---   &4.21(-15)&1.22(-14)&7.54(-14)&5.09(-12)&2.43(-11)&7.56(-11)     &7,11,14,15,16\\
GM Cha             &+&   ---   &   ---   &   ---   &   ---   &1.54(-12)&9.09(-12)&3.79(-11)     &  7     \\
                   &-&   ---   &   ---   &   ---   &   ---   &2.09(-13)&1.43(-12)&7.70(-12)     &  17    \\
V2492 Cyg$^e$      &+&   ---   &   ---   &   ---   &   ---   &2.25(-10)&3.85(-10)&5.28(-10)     &  18    \\
                   &-&   ---   &   ---   &   ---   &   ---   &7.31(-14)&4.51(-13)&5.18(-12)     &  19    \\
V2493 Cyg$^f$      &+&   ---   &   ---   &1.33(-10)&1.86(-10)&3.49(-10)&4.19(-10)&3.27(-10)     &  20    \\
                   &-&   ---   &   ---   &2.41(-12)&7.62(-12)&1.94(-11)&2.41(-11)&2.41(-11)     &  7,21  \\
\hline

\enddata
\tablenotetext{a}{+ = high state; - = low state}
\tablenotetext{b}{photometric bands are conventionally associated (see text) to the following effective wavelengths (in $\mu$m):
B 0.43, V 0.55, R 0.64, I 0.79, J 1.25, H 1.65, K 2.20}
\tablenotetext{c}{Fluxes are also obtained at U (0.36$\mu$m: 1.32(-9)/3.08(-11); M (4.80$\mu$m): 3.20(-9)/1.10(-10).}
\tablenotetext{d}{Fluxes are also obtained at L' (3.76$\mu$m): 1.56(-9)/1.06(-9).}
\tablenotetext{e}{Fluxes are also obtained at L' (3.75$\mu$m): 1.43(-9)/6.16(-11).}
\tablenotetext{f}{Fluxes are also obtained at $z'$ (0.90$\mu$m): 2.65(-10)/6.06(-12).}

\tablecomments{References to the Table: (1) Audard et al. 2010; (2) Lorenzetti et al. 2006; (3) Aspin et al. 2010;
(4)Sipos et al. 2009; (5) Lorenzetti et al. 2011;  (6) Kun et al. 2011b; (7) Cutri et al. 2003 (2MASS); (8) Liseau et al. 1992;
(9) Eisl\"{o}ffel et al. 1991; (10) Caratti o Garatti et al. 2011; (11) Brice\~{n}o et al. 2004; (12) Reipurth \& Aspin 2004;
(13) Andrews et al. 2004; (14)  K\'{o}sp\'{a}l et al. 2005;  (15) Aspin \& Reipurth 2009; (16) \'{A}brah\'{a}m et al. 2004; (17) Persi et al. 2007; (18) Covey et al. 2011; (19) Aspin 2011; (20) Miller et al. 2011; (21) Guieu et al. 2009.  }

\end{deluxetable}

\begin{deluxetable}{lccccc}
\tabletypesize{\footnotesize}
\tablewidth{0pt}
\tablecaption{Ranges of the fit parameters corresponding to the best $\chi^2$ values compatible with the spot model (blue and green regions in Figure~\ref{maps:fig}). \label{fit:tab}}
\tablehead{Source & $\chi^2$ & T (K) & R (AU) & A$^{quiesc}_V$ (mag) & $\Delta A_V$ (mag) }
\startdata
V1118Ori   & 1.64 & 4546$^{+1454}_{-460}$    & 0.01$^{+0.00}_{-0.00}$ & 2.1$^{+2.0}_{-2.1}$  & -1.9$^{+2.4}_{-0.7}$ \\
           &                                 &                        &                      &                      \\
EXLup      & 21.7 & 20332$^{+14668}_{-19242}$& 0.002$^{+0.12}_{-0.00}$& 3.1$^{+11.9}_{-0.9}$ & -1.5$^{+0.4}_{-1.4}$ \\
           &                                 &                        &                      &                      \\
PVCep      & 4.37 & 1554$^{+211}_{-82}$      & 0.14$^{+0.02}_{-0.02}$ & 0.9$^{+1.8}_{-0.6}$  & -0.7$^{+0.4}_{-0.3}$ \\
           &                                 &                        &                      &                      \\
\hline
           &                                 &                        &                      &                      \\
V1180 Cas  &15.54 & 1981$^{+67}_{-98}$       & 0.03$^{+0.00}_{-0.00}$ & 4.3$^{+0.9}_{-0.4}$  & -4.2$^{+0.3}_{-0.7}$ \\
           &                                 &                        &                      &                      \\
SVS13      & 4.60 & 1468$^{+685}_{-407}$     & 0.09$^{+0.08}_{-0.04}$ & 9.8$^{+5.2}_{-6.5}$  & -3.4$^{+0.3}_{-0.1}$ \\
           &                                 &                        &                      &                      \\
V2775 Ori  & 0.30 & 1212$^{+3162}_{-46}$     & 0.25$^{+0.02}_{-0.18}$ & 0.0$^{+15}_{-0.0}$   & -0.1$^{+4.9}_{-5.1}$ \\
           &                                 &                        &                      &                      \\
V1647Ori   &29.06 & 1554$^{+82}_{-61}$       & 0.13$^{+0.01}_{-0.01}$ & 5.5$^{+0.2}_{-3.4}$  & -5.0$^{+3.0}_{-0.0}$ \\
           &                                 &                        &                      &                      \\
GM Cha     & 0.70 & 870$^{+227}_{-22}$       & 0.06$^{+0.01}_{-0.02}$ & 5.8$^{+8.8}_{-2.7}$  & -5.0$^{+1.9}_{-0.0}$ \\
           &                                 &                        &                      &                      \\
V2492Cyg   & 12.6 & 1383$^{+178}_{-70}$      & 0.27$^{+0.01}_{-0.06}$ & 4.6$^{+2.2}_{-4.6}$  & -4.6$^{+4.6}_{-0.4}$ \\
           &                                 &                        &                      &                      \\
V2493 Cyg  & 18.07 & 2580$^{+183}_{-17}$     & 0.06$^{+0.00}_{-0.01}$ & 4.3$^{+0.3}_{-4.3}$  & -4.2$^{+4.5}_{-0.2}$ \\
           &                                 &                        &                      &                      \\
\enddata
\end{deluxetable} 
\begin{deluxetable}{lccccccc}
\tabletypesize{\footnotesize} \tablecaption{Optical and near-IR photometry of V2493 Cyg. \label{HBC722_data:tab}}
\tablewidth{0pt}
\tablehead{ Date  &   B   &   V   &   R   &   I   &  J &  H  &  K   \\
   JD-2400000     &  \multicolumn{7}{c}{(mag)}   }
\startdata
43449$^a$         &  ---  &  ---  &  ---  &  ---  &  ---  & 11.12 & 10.26 \\
51705$^b$         &  ---  &  ---  &  ---  &  ---  & 13.25 & 12.21 & 11.46 \\
\hline
55440$^c$         &  ---  &  ---  &  ---  &  ---  & 10.24 &  9.27 &  8.74 \\
55497             &  ---  & 13.98 & 12.85 & 11.86 &  ---  &  ---  &  ---  \\
55500$^c$         &  ---  &  ---  &  ---  &  ---  & 10.29 &  9.30 &  8.76 \\
55504             &  ---  &  ---  &  ---  &  ---  & 10.25 &  9.21 &  8.76 \\
55507             &  ---  &  ---  &  ---  &  ---  & 10.36 &  9.25 &  8.70 \\
55513             &  ---  &  ---  &  ---  &  ---  & 10.27 &  9.36 &  8.80 \\
55514             &  ---  &  ---  &  ---  &  ---  & 10.39 &  9.37 &  8.80 \\
55515             &  ---  &  ---  &  ---  &  ---  & 10.43 &  9.39 &  8.82 \\
55516             &  ---  &  ---  &  ---  &  ---  & 10.41 &  9.42 &  8.82 \\
55561             & 15.98 & 14.44 & 13.42 & 12.33 &  ---  &  ---  &  ---  \\
55577             &  ---  & 14.68 & 13.63 & 12.51 &  ---  &  ---  &  ---  \\
55588             & 16.22 & 14.81 & 13.77 & 12.57 &  ---  &  ---  &  ---  \\
55673             & 16.64 & 15.11 &  ---  &  ---  &  ---  &  ---  &  ---  \\
55678             &  ---  &  ---  & 13.89 & 12.63 &  ---  &  ---  &  ---  \\
55689             & 16.71 & 15.10 &  ---  & 12.75 &  ---  &  ---  &  ---  \\
55732             & 16.70 & 15.08 & 13.88 & 12.60 &  ---  &  ---  &  ---  \\
55741             & 16.53 & 14.92 & 13.78 & 12.52 &  ---  &  ---  &  ---  \\
55746             & 16.62 & 14.97 & 13.79 & 12.57 &  ---  &  ---  &  ---  \\
55755             & 16.53 & 14.93 & 13.76 & 12.53 &  ---  &  ---  &  ---  \\
55759             & 16.52 & 14.97 & 13.79 & 12.53 &  ---  &  ---  &  ---  \\
55773             & 16.59 & 15.11 & 13.88 & 12.61 &  ---  &  ---  &  ---  \\
55775             & 16.61 & 15.08 &  ---  & 12.54 &  ---  &  ---  &  ---  \\
55778$^c$         &  ---  &  ---  &  ---  &  ---  & 10.70 &  9.61 &  9.01 \\
55783             & 16.64 & 15.05 & 13.85 & 12.59 &  ---  &  ---  &  ---  \\
55784             &  ---  &  ---  &  ---  &  ---  & 10.68 &  9.59 &  8.97 \\
55786             & 16.89 & 15.03 & 13.85 & 12.58 & 10.61 &  9.54 &  8.97 \\
55787             &  ---  &  ---  &  ---  &  ---  & 10.61 &  9.55 &  8.98 \\
55789             &  ---  &  ---  &  ---  &  ---  & 10.68 &  9.61 &  8.99 \\
55790             &  ---  &  ---  &  ---  &  ---  & 10.72 &  9.63 &  9.04 \\
55791             &  ---  &  ---  &  ---  &  ---  & 10.67 &  9.57 &  8.99 \\
55792             &  ---  &  ---  &  ---  &  ---  & 10.63 &  9.59 &  8.98 \\
55793             &  ---  &  ---  &  ---  &  ---  & 10.63 &  9.57 &  8.97 \\
55794             &  ---  &  ---  &  ---  &  ---  & 10.64 &  9.55 &  8.97 \\
55796             &  ---  &  ---  &  ---  &  ---  & 10.65 &  9.58 &  8.97 \\
55797$^c$         &  ---  &  ---  &  ---  &  ---  & 10.59 &  9.56 &  8.95 \\
55798             & 16.60 & 14.91 & 13.89 & 12.57 & 10.58 &  9.55 &  8.95 \\
55799             &  ---  &  ---  &  ---  &  ---  & 10.61 &  9.53 &  8.95 \\
55800             & 16.60 & 14.99 & 13.85 & 12.56 &  ---  &  ---  &  ---  \\
55802             & 16.51 & 14.93 & 13.75 & 12.52 &  ---  &  ---  &  ---  \\
55805             &  ---  &  ---  &  ---  &  ---  & 10.69 &  9.61 &  9.00 \\
55808$^c$         &  ---  &  ---  &  ---  &  ---  & 10.67 &  9.61 &  9.02 \\
55811             & 16.67 &  ---  & 13.92 & 12.63 &  ---  &  ---  &  ---  \\
55815             &  ---  &  ---  & 13.88 & 12.62 &  ---  &  ---  &  ---  \\
55817             &  ---  & 15.12 & 13.90 & 12.61 &  ---  &  ---  &  ---  \\

\enddata
\tablenotetext{a}{Early observation (1977) by CoKu.}
\tablenotetext{b}{2MASS observation (2000).}
\tablenotetext{c}{Date on which a near-IR spectrum has been taken (see Figure~\ref{spectrum:fig}).}

\tablecomments{Typical errors of the optical/near-IR magnitude are less than 0.04 mag.}

\end{deluxetable}


\newpage
\normalsize
\begin{thebibliography}{}
\bibitem{} \'{A}brah\'{a}m, P., K\'{o}sp\'{a}l, \'{A}., Csizmadia, Sz., Mo\'{o}r, A., Kun,M., \& Strigfellow, G.  2004, A\&A, 419, L39
\bibitem{} Alves de Oliveira, C., \& Casali, M. 2008, A\&A, 485, 155A
\bibitem{} Andrews, S.M., Rothberg, B., \& Simon, T. 2004, ApJ, 610, L45
\bibitem{} Antoniucci, S. et al. 2011, A\&A, 534A, 32A
\bibitem{} Aspin, C. 2011, AJ, 141:196
\bibitem{} Aspin, C., Beck, T.L., \& Reipurt, B. 2008, AJ, 135, 423
\bibitem{} Aspin, C., \& Reipurt, B. 2009, AJ, 138, 1137A
\bibitem{} Aspin, C., Reipurt, B., Herczeg, G.J., \& Capak, P. 2010, ApJ Lett., 719, L50
\bibitem{} Audard, M., Stringfellow, G.S., G\"{u}del, M. et al. 2010 A\&A 511, 63
\bibitem{} Back, T.L., Bary, J.S. \& McGregor, P.J., 2010, ApJ 722, 1360
\bibitem{} Bloom, J.S., Starr, D.L., Blake, C.H., Skrutskie, M.F., \& Falco, E.E. 2006, in Astronomical Society
of the Pacific Conference Series, Vol. 351, ed. C.Gabriel, C.Arvsiet, D.Ponz \& S.Enrique, 751
\bibitem{} Breger, M., Gehrz, R.D., \& Hackwell, J.A. 1981, ApJ, 248, 963
\bibitem{} Brice\~{n}o, C., et al. 2004, ApJ, 606, L123
\bibitem{} Calvet, N., Hartmann, L., \& Strom, S.E. 2000, Protostars and Planets IV - University of Arizona Press;
eds V.Mannings, A.P.Boss, S.S Russell, p.377
\bibitem{} Calvet, N., \& Gullbring, E. 1998, ApJ, 509, 802
\bibitem{} Cardelli, J.A., Clayton, G.C. \& Mathis, J.S. 1989, ApJ, 345, 245
\bibitem{} Caratti o Garatti, A. et al. 2011, A\&A, 526, L1
\bibitem{} Cohen, M., \& Kuhi, L. 1979, ApJS, 41, 743
\bibitem{} Cohen, M., Kuhi, L.V., Harlan, E.A., \& Spinrad, H. 1981, ApJ, 245, 920
\bibitem{} Covey, K.R. et al. 2011, AJ, 141, 40
\bibitem{} Cutri, R. M., et al. 2003, Explanatory Supplement to the 2MASS All Sky Data
Release(Pasadena: Caltech)
\bibitem{} D'Angelo, C.R., \& Spruit, H.C. 2010 MNRAS 406, 1208
\bibitem{} Davis, C.J. et al. 2011, A\&A, 528, A3
\bibitem{} Eisl\~{o}ffel, J., Eike, G., Hessman, F.V., Mundt, R., Poetzen, R., Carr, J.S.,
Beckwith, S., \& Ray, T.P. 1991, ApJ, 383, L19
\bibitem{} Fernie, J.D. 1983 PASP 95, 782
\bibitem{} Fischer, W., Edwards, S., Hillenbrand, L., \& Kwan, J. 2011 ApJ 730-73
\bibitem{} Gail, H.-P., \& Sedlmayr, E. 1999 A\&A 347, 594
\bibitem{} Giannini, T. et al. 2009, ApJ, 704, 606
\bibitem{} Giovanardi, C., Gennari, S., Natta, A., \& Stanga, R. 1991 ApJ 367, 173
\bibitem{} G\'{o}mez, M., \& Mardones, D. 2003 AJ 125, 2134
\bibitem{} Gras-Vel\'{a}zquez, A., \& Ray, T.P. 2005, A\&A, 443, 541
\bibitem{} Guieu, S. et al. 2009 ApJ 697, 787
\bibitem{} Gullbring, E., Hartmann, L., Bri\~{c}eno, C., \& Calvet, N. 1998, ApJ, 492, 323
\bibitem{} Hartigan, P., Edwards, S.,\& Ghandour, L. 1995, ApJ, 452, 736
\bibitem{} Hartmann, L., \& Kenyon, S. 1985, ApJ, 299, 462
\bibitem{} Herbig, G.H. 1989, Proc. of the ESO Workshop on {\it Low Mass Star Formation and
Pre-Main Sequence Objects}, ed. B. Reipurth, p.233
\bibitem{} Herbig, G.H. 1990, ApJ, 360, 639
\bibitem{} Herbig, G.H. 2008, AJ, 135, 637
\bibitem{} Herbig, G.H., Aspin, C., Gilmore, A.C., Imhoff, C.L., Jones, A.F. 2001, PASP, 113, 1547
\bibitem{} Herbig, G.H., \& Bell, K.R. 1988, Lick Obs. Bulletin No.1111
\bibitem{} Herbst, W., Herbst, D.K., Grossman, E.J., \& Weinstein, D. 1994, AJ, 108, 1906
\bibitem{} Juh\'{a}sz, A. et al. 2011, arXiv: 1110, 3754
\bibitem{} Kazarovets, E.V., Reipurth, B., \& Samus, N.N. 2011, Variablen Stars, 31, vol.2
\bibitem{} Kenyon, S.J. et al. 1994, AJ, 107, 2153
\bibitem{} K\'{o}sp\'{a}l, \'{A}., \'{A}brah\'{a}m, P., Prusti, T., Acosta-Pulido, J., Hony, S.,
Mo\'{o}r, A., \& Siebenmorgen, R. 2007, A\&A, 470, 211
\bibitem{} K\'{o}sp\'{a}l, \'{A}. et al. 2005, IBVS, 5661
\bibitem{} K\'{o}sp\'{a}l, \'{A}. et al. 2011, A\&A, 527, A133
\bibitem{} Kun, M. et al. 2011a, MNRAS, 413, 2689
\bibitem{} Kun, M. et al. 2011b, ApJL, 733, L8
\bibitem{} Leoni, R., Larionov, V.M., Centrone, M., Giannini, T., \& Lorenzetti, D. 2010 Astronomer's Telegram \#2854
\bibitem{} Liseau, R., Lorenzetti, D., \& Molinari, S. 1992, A\&A, 253, 119
\bibitem{} Lorenzetti, D. et al. 2006, A\&A, 453, 579
\bibitem{} Lorenzetti, D. et al. 2011b, ApJ, 723:69
\bibitem{} Lorenzetti, D., Giannini, T., Larionov, V.M., Kopatskaya, E., Arkharov, A.A.,
De Luca, M., \& Di Paola, A. 2007, ApJ, 665, 1193
\bibitem{} Lorenzetti, D., Larionov, V.M., Giannini, T., Arkharov, A.A., Antoniucci, S.,
Nisini, B., \& Di Paola, A. 2009, ApJ, 693, 1056
\bibitem{} McGehee, P.M. et al. 2004, ApJ, 616, 1058
\bibitem{} Menten, K.M., Reid, M.J., Forbrich, J., \& Brunthaler, A. 2007, A\&A, 474, 515
\bibitem{} Meyer, M.R., Calvet, N. \& Hillenbrand, L.A. 1997, AJ, 114, 288
\bibitem{} Miller, A.A. et al. 2011, ApJ, 730, 80
\bibitem{} Muzerolle, J., Calvet, N., Hartmann, L., \& D'Alessio, P. 2003, ApJ, 597, 149
\bibitem{} Muzerolle, J., D'Alessio, P., Calvet, N., \&  Hartmann, L. 2004, ApJ, 617, 406
\bibitem{} Muzerolle, J., Megeath, S.T., Flaherty, K.M., Gordon, K.D., Rieke, G.H., Young, E.T., \& Lada, C.J. 2005, ApJ, 620, L107
\bibitem{} Muzerolle, J., Hartmann, L., \& Calvet, N. 1998, AJ, 116, 2965
\bibitem{} Meyer, M.R., Calvet, N. \& Hillenbrand, L.A. 1997, AJ, 114, 288
\bibitem{} Persi, P., Tapia, M., G\`{o}mez, M., Whitney, B.A., Marenzi, A.M., \& Roth, M. 2007, AJ, 133, 1690
\bibitem{} Reipurth, B., \& Aspin, C. 2004, ApJ 606, L119
\bibitem{} Rieke, G.H. \& Lebofsky, M.J. 1985, ApJ, 288, 618
\bibitem{} Semkov, E., \& Peneva, S. 2010, Astronomer's Telegram \#2801
\bibitem{} Sharp, C.M., \& Huebner, W.F. 1990, ApJ, 72, 417
\bibitem{} Sicilia-Aguilar, A. et al. 2008, ApJ, 673, 382
\bibitem{} Shu, F.H., Najita, J.R., Ostriker, E., Wilkin, F., Ruden, S.,
\& Lizano, S. 1994, ApJ, 429, 781
\bibitem{} Sipos, N. et al. 2009, A\&A, 507, 881S
\bibitem{} Stecklum, B., Melnikov, S.Y., \& Meusinger, H. 2007, A\&A, 175, 231
\bibitem{} Straizys, V., Cernis, K., Kazlauskas, A., \& Meistas, E. 1992 Balt.Astr., 1, 149
\bibitem{} Takami, M. et al. 2006, ApJ, 641, 357
\bibitem{} Van Citters Jr., G.W., \& Smith, S. 1989, AJ, 98, 1328
\bibitem{} Walter, F.M., Stringfellow, G.S., Sherry, W.H., \& Field-Pollatou, A. 2004, AJ, 128, 1872
\bibitem{} Xiao, L., Kroll, P., \& Henden, A.A. 2010, AJ, 139, 1527
\end{thebibliography}
\end{document}